\documentclass[letterpaper,twocolumn,10pt]{article}
\usepackage{usenix2019_v3}
\usepackage{amsmath}
\usepackage{amsfonts}              
\usepackage{amssymb}               
\usepackage{tabularx}
\usepackage{comment}
\usepackage{gensymb}
\usepackage{caption}
\usepackage{subcaption}

\usepackage{graphicx}
\DeclareGraphicsExtensions{.pdf,.jpeg,.png}

\usepackage{algpseudocode,algorithm,algorithmicx}


\usepackage{array}
\newcolumntype{C}[1]{>{\centering\let\newline\\\arraybackslash\hspace{0pt}}m{#1}}
\usepackage[normalem]{ulem}

\usepackage{cite}

\usepackage{mathptmx}              
\usepackage{helvet}        
\usepackage{inconsolata}              
\usepackage{textcomp}              

\usepackage{pifont} 
\usepackage{enumitem}

\usepackage{booktabs}              

\usepackage{relsize}               
\usepackage{xspace}                
\usepackage[all]{nowidow}
\usepackage[svgnames,table]{xcolor}      

\usepackage{colortbl}
\usepackage{multirow}
\usepackage{hyphenat}
\usepackage{balance}  

\usepackage{hyperref}
\usepackage{cleveref}

\crefformat{section}{\S#2#1#3} 
\crefformat{subsection}{\S#2#1#3}
\crefformat{subsubsection}{\S#2#1#3}
\Crefname{section}{\S}{\S}
\Crefname{subsection}{\S}{\S}
\Crefname{subsubsection}{\S}{\S}
\crefname{figure}{Figure}{Figures}
\Crefname{figure}{Figure}{Figures}
\crefname{subfigure}{Figure}{Figures}
\Crefname{subfigure}{Figure}{Figures}
\crefrangelabelformat{section}{#3#1#4--#5#2#6}
\crefrangelabelformat{subsection}{#3#1#4--#5#2#6}

\crefrangelabelformat{subfigure}{#3#1#4--#5(\crefstripprefix{#1}{#2}#6}

\usepackage{listings}              
\lstdefinestyle{VRQL}{
    language=C++,
    basicstyle=\scriptsize\ttfamily,
    mathescape=true,
    xleftmargin=0.25cm,
    columns=fullflexible,
    morekeywords={Union,Decode,Scan,Select,Map,Partition,Encode,Subquery,Interpolate,Translate,Rotate,Discretize,Store,Update,Create,CreateIndex,DropIndex}
 }

\newcommand{\name}{Vignette\xspace}
\newcommand{\nameStore}{Vignette Storage\xspace}
\newcommand{\nameCompress}{Vignette Compression\xspace}

\newcommand{\etal}{et al.}

\newcommand{\avc}{\textsc{h.264}\xspace}
\newcommand{\hevc}{\textsc{hevc}\xspace}

\newcommand{\avone}{\textsc{av1}\xspace}
\newcommand{\threesixty}{360\degree\xspace}
\newcommand{\lightdb}{LightDB\xspace}

\newcommand{\tileVsSizeFigure}{
\begin{figure}[t]
  \centering
\includegraphics[width=\linewidth]{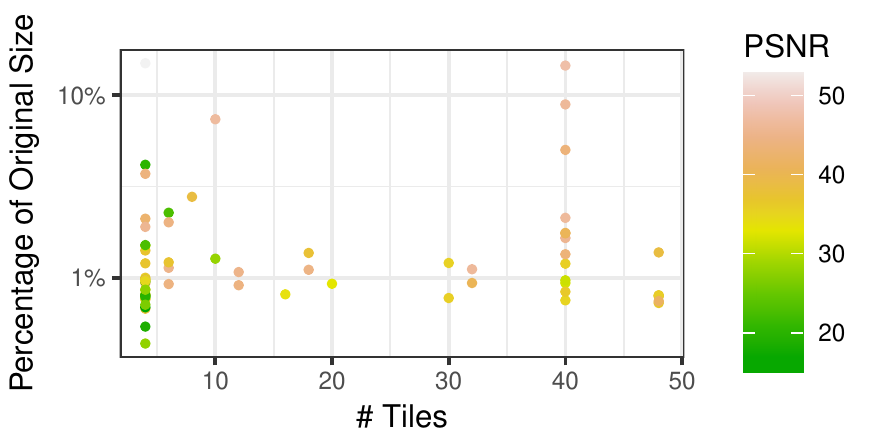}
  \caption{Compression ratio and PSNR at the optimal number of tiles for each video. The optimal number of tiles is video-dependent, not correlated with quality or compression ratio.}
  \label{plot:tile-vs-size}
\end{figure}
}

\newcommand{\computeTable}{
\begin{table}[tb]
\centering
\caption{Mean processing time per video, evaluated over all videos in our datasets.}
\label{table:compute}
\resizebox{\linewidth}{!}{%

\begin{tabular}{lrcrc} \toprule
   & \multicolumn{2}{c}{Exhaustive} & \multicolumn{2}{c}{Heuristic }\\
    \cmidrule(lr){2-3}\cmidrule(lr){4-5}
  Task                       & Time (s) & \%   & Time (s) & \% \\ \midrule
  Generate saliency map      & 1633     & 49\% &  1633     & 95\%   \\
  Compute tile configuration & 1696     & 50\phantom{\%} &  59      & 4\phantom{\%}   \\
  Saliency-based transcode   & 21       & \phantom{0}1\phantom{\%}  &  21      & \phantom{0}1\phantom{\%}   \\ \midrule
  Total                      & 3350     &      &  1713     & \\ \bottomrule
\end{tabular}
}
\end{table}

}

\newcommand{\userStudyFigure}{

\begin{figure}[t]
  \centering
  \includegraphics[width=\linewidth]{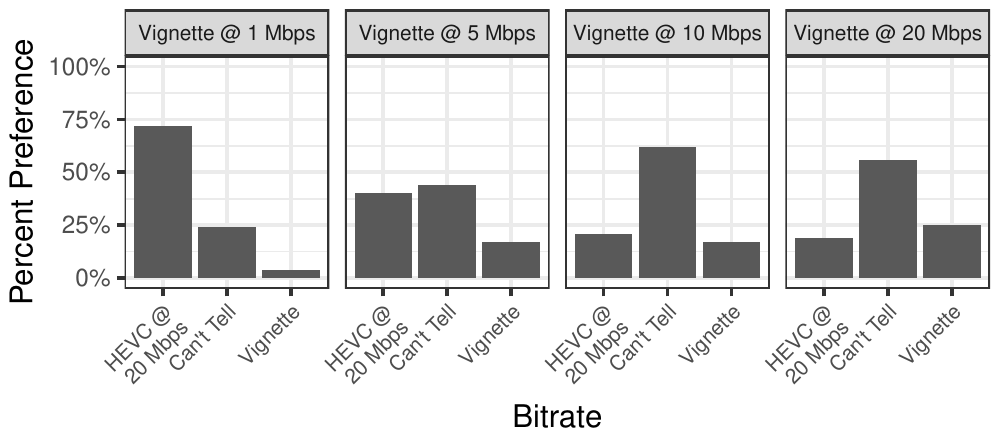}
  \caption{Results of perceived quality preference user study, averaged across participants and videos by bitrate. Participants either preferred \name or perceived no difference between 20 Mbps \hevc videos and \name videos at 5--20 Mbps.}
  \label{plot:user-study}
\end{figure}
}

\newcommand{\saliencyTilesOverviewFigure}{
\begin{figure}[t]
  \centering
  \includegraphics[width=\linewidth]{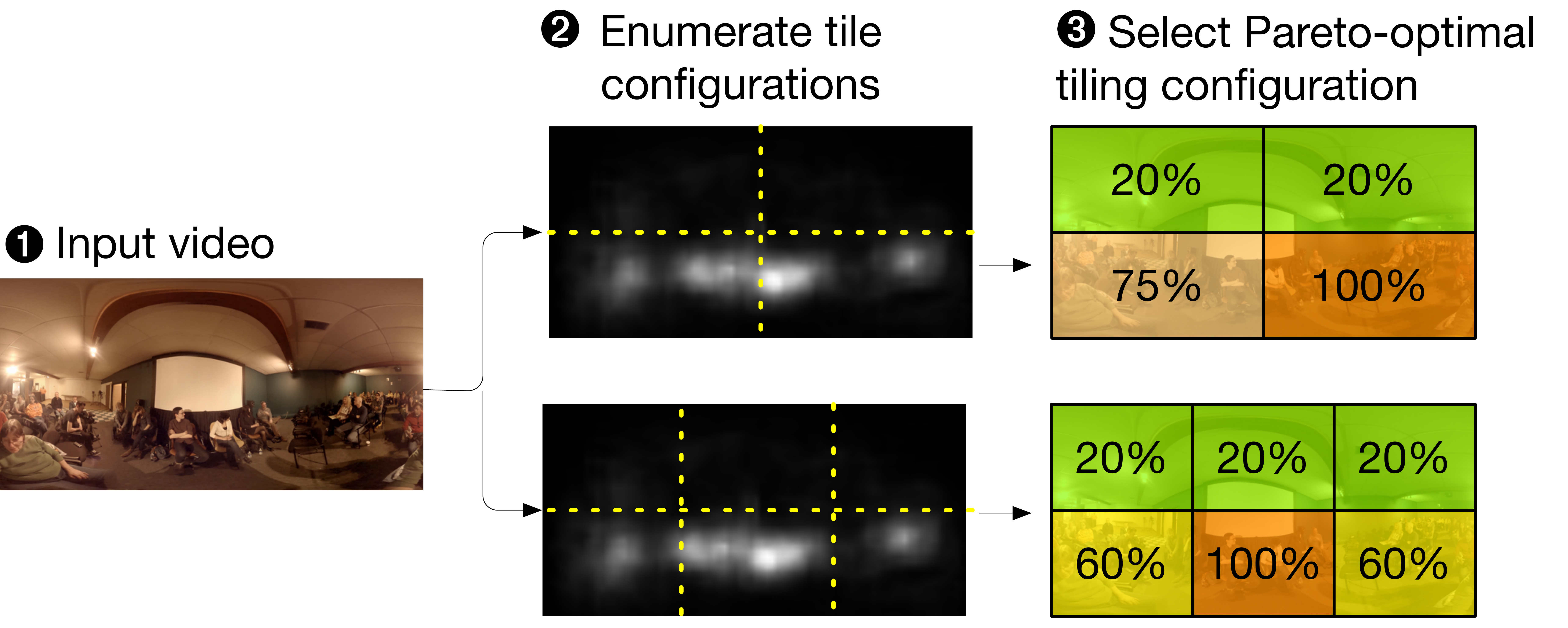}
  \caption{Overview of \nameCompress algorithm.}
  \label{fig:saliency-tiles-overview}
\end{figure}
}

\newcommand{\architectureOverviewFigure}{
\begin{figure}[t]
  \centering
  \includegraphics[width=.9\linewidth]{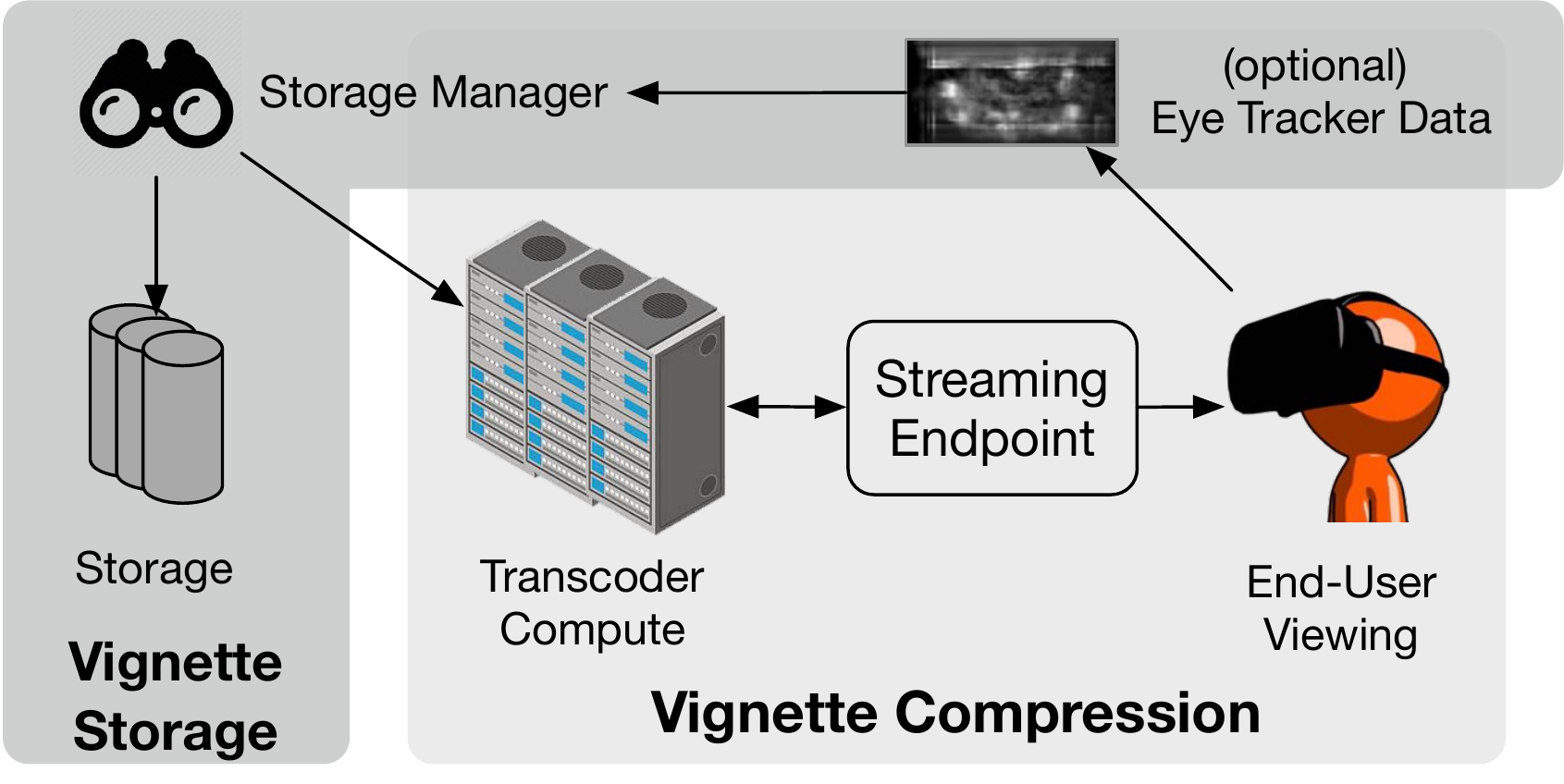}
  \caption{\name provides two features: \nameCompress, a perceptual compression algorithm, and \nameStore, a storage manager for perceptually compressed videos.
  Integrating perceptual information with the storage manager reduces network bandwidth and storage costs.}
  \label{fig:high-level-architecture}
\end{figure}
}

\newcommand{\exampleVidSaliencyFigure}{

\begin{figure*}[ht]
  \centering
  \begin{subfigure}{.3\textwidth}
    \centering
    \includegraphics[width=.9\linewidth]{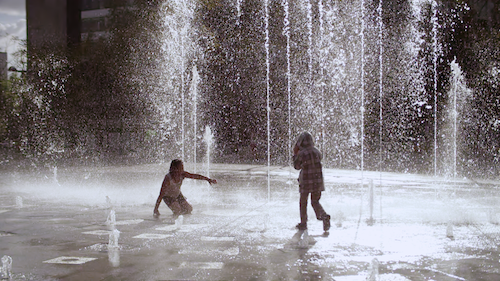}
    \caption{Input video frame from Netflix~\cite{netflix2016data}.}
    \label{subfig:og-stil}
  \end{subfigure}%
  \begin{subfigure}{.3\textwidth}
    \centering
    \includegraphics[width=.9\linewidth]{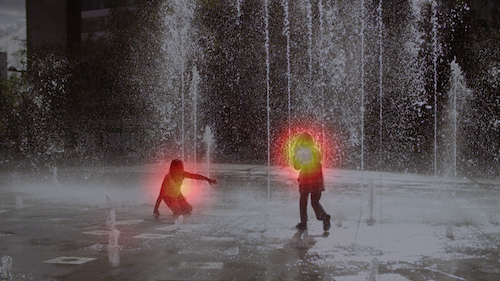}
    \caption{Saliency map produced by MLNet~\cite{mlnet2016}, overlaid on input.}
    \label{subfig:sal-still}
  \end{subfigure}%
  \begin{subfigure}{.3\textwidth}
    \centering
    \includegraphics[width=.9\linewidth]{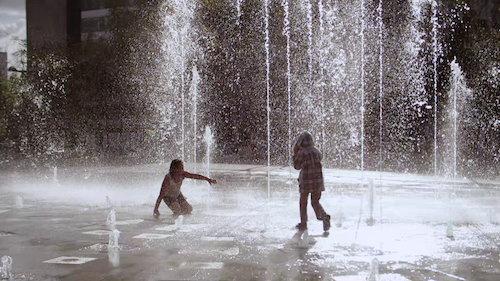}
    \caption{Perceptually-compressed \name video, 85\% smaller at iso-quality.}
    \label{subfig:sal-vign}
  \end{subfigure}%

  \caption{Example video still, neural network-generated saliency map, and output \name perceptually compressed video.}
  \label{fig:example-vid-saliency}
\end{figure*}

}

\newcommand{\systemOverviewFigure}{
\begin{figure*}[ht]
\begin{subfigure}{.48\textwidth}
  \centering
  \includegraphics[width=.9\linewidth]{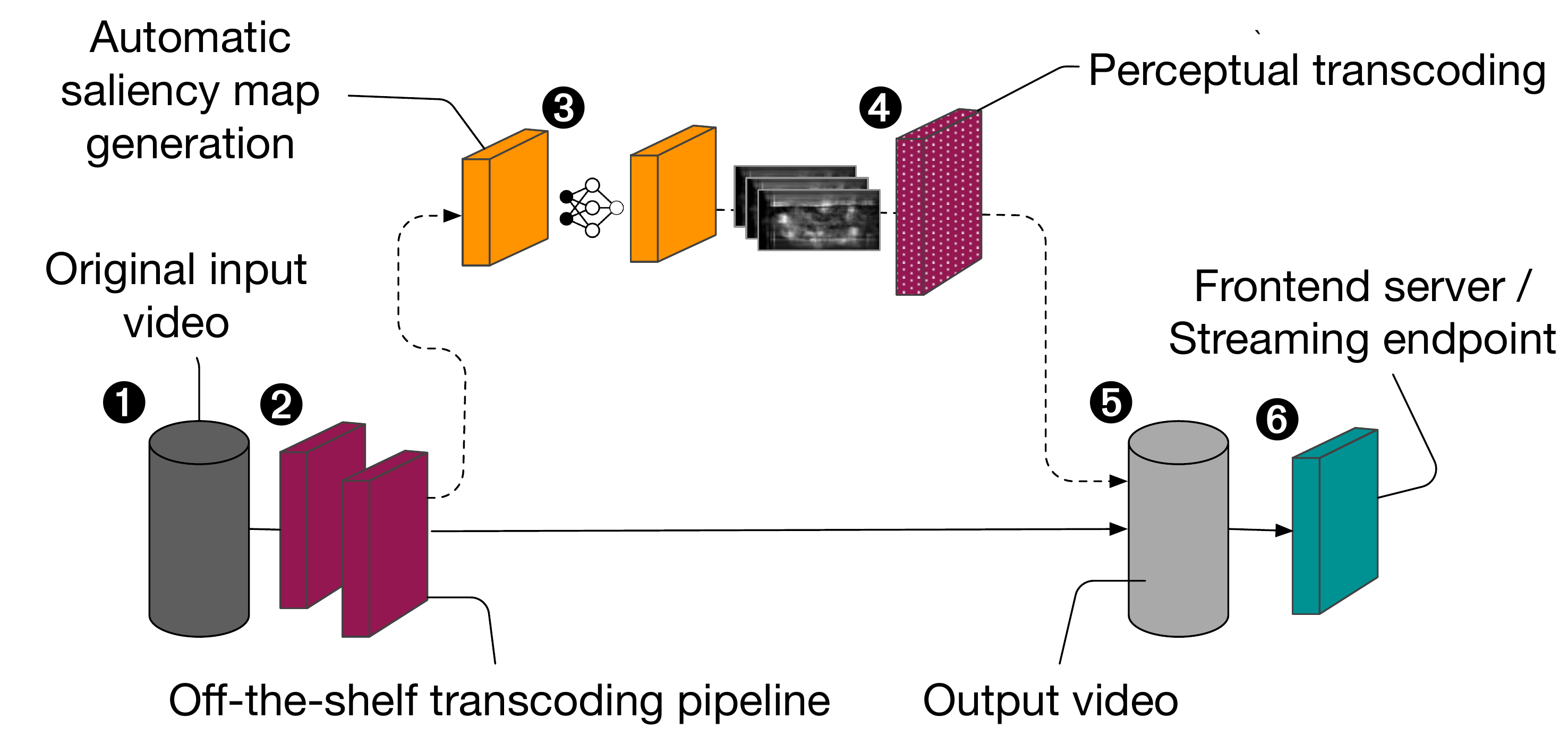}
  \caption{Open-loop offline saliency compression:~\name automatically generates saliency maps to include perceptual data during video transcoding.}
  \label{fig:system-overview-open}
\end{subfigure}\hfill%
\begin{subfigure}{.48\textwidth}
  \centering
  \includegraphics[width=.9\linewidth]{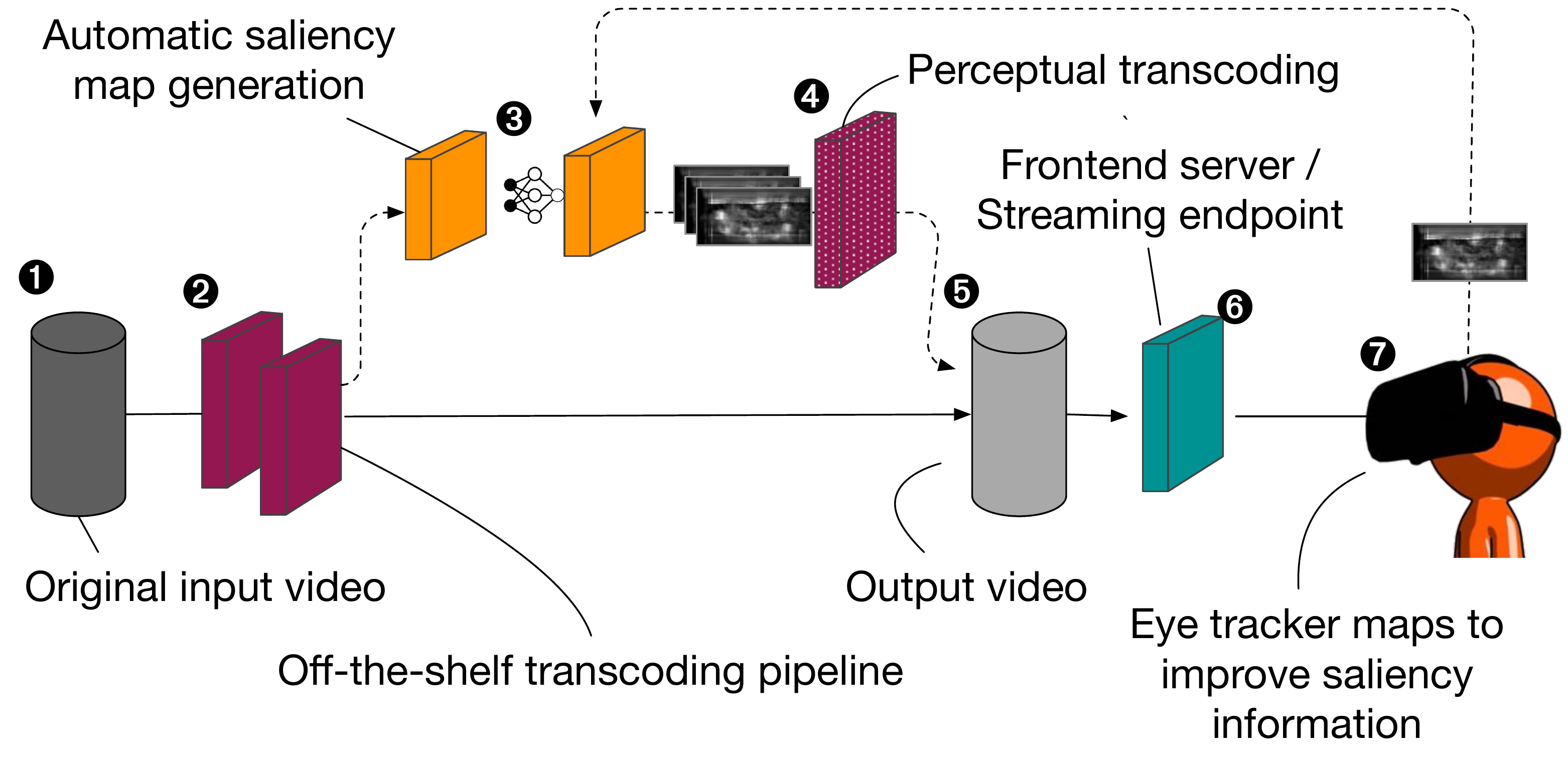}
  \caption{Closed-loop re-compression:~\name~can leverage perceptual cues from VR headsets and other eyetracking devices to improve compression.}
  \label{fig:sub2}
\end{subfigure}
\caption{High-level architecture of~\name~system design.}
  \label{fig:system-overview}
\end{figure*}
}

\newcommand{\videoMetadataFigure}{
\begin{figure}[t]
  \centering
  \includegraphics[width=.95\linewidth]{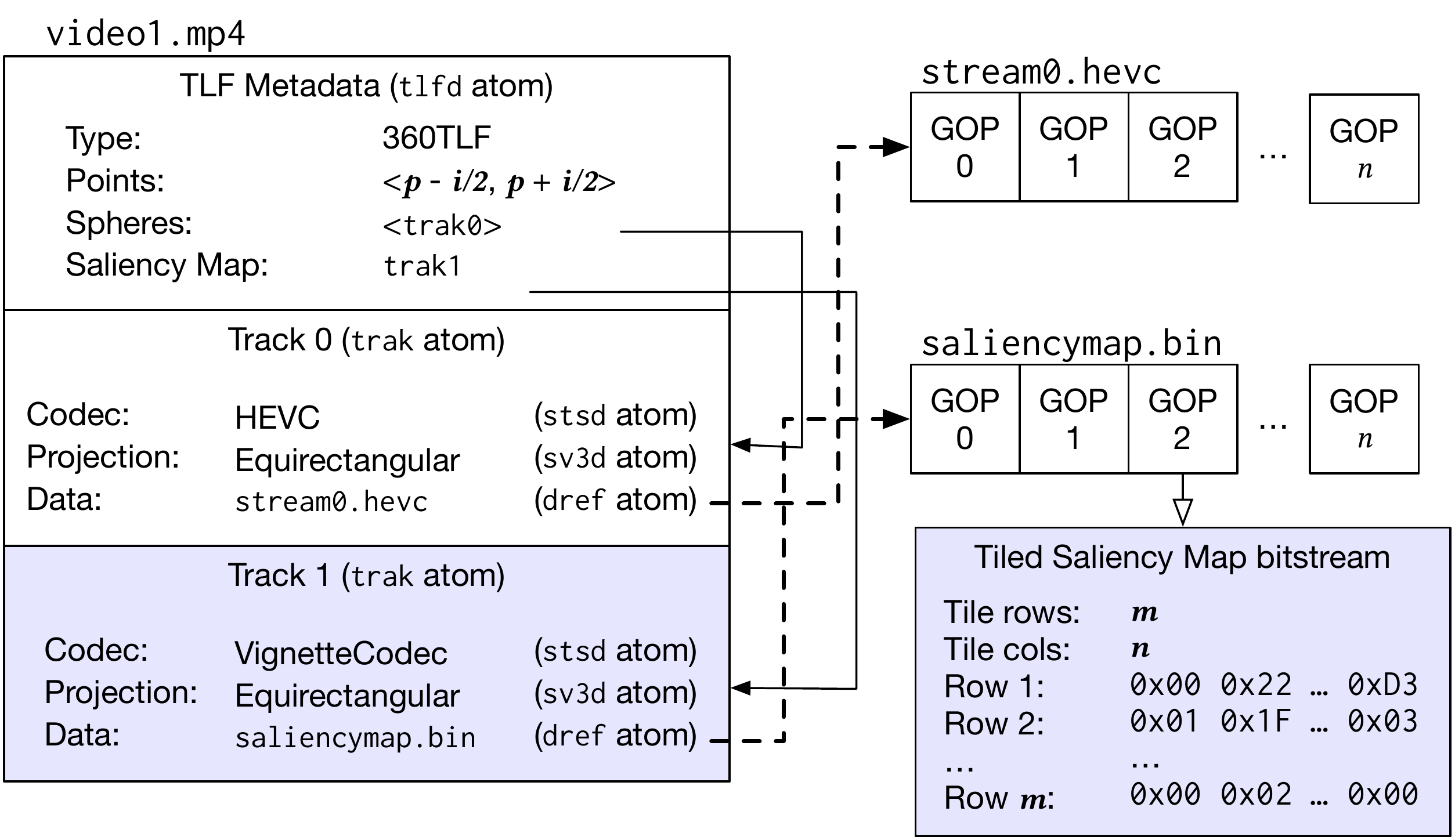}
  \caption{Physical layout of video metadata in~\lightdb. \name-specific features are highlighted.}
  \label{fig:video-metadata}
\end{figure}
}

\newcommand{\benchmarkInformationFigure}{
\begin{table}[t]
  \centering
       \resizebox{\linewidth}{!}{%
       \begin{tabular}{l l l l l} \toprule
       Type                      & Benchmark & Description & Bitrate (Mbps) & Size (MB)        \\\midrule
       \multirow{2}{*}{Standard} & vbench~\cite{vbench}    & YouTube dataset & 0.53--470 & 757  \\
                                 & Netflix~\cite{netflix2016data}   & Netflix dataset & 52--267 & 1123 \\
       \multirow{2}{*}{VR}       & VR-360~\cite{saliency-map}    & 4K-360 dataset & 10--21 & 1400   \\
                                 & Blender~\cite{blender}   & UHD / 3D movies & 10--147 & 6817 \\ \bottomrule
       \end{tabular}
       }

  \caption{Video datasets used to characterize \name. }
  \label{table:benchmarks}
\end{table}

}

\newcommand{\avgStorageBitrateSavings}{
\begin{figure}[t]
  \centering
    \includegraphics[width=\linewidth]{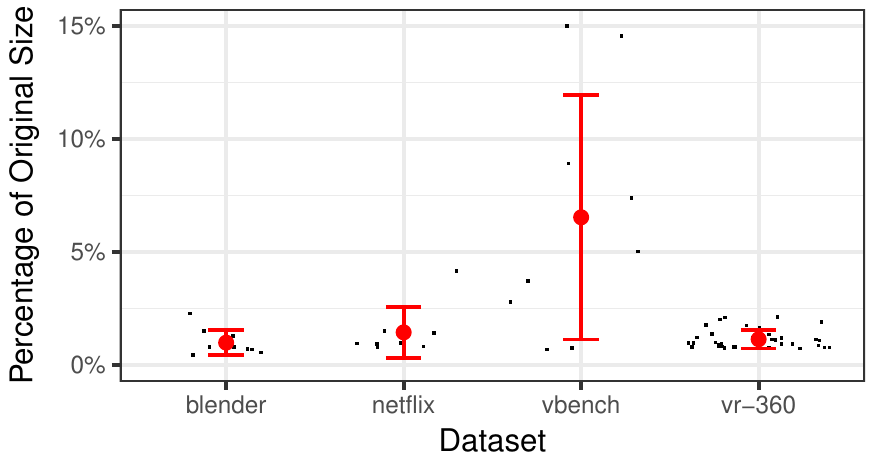}

  \caption{Aggregate storage savings by dataset. \nameCompress reduces videos to 1--15\% of their original size while maintaining PSNR of 34--39 dB and EWPSNR of 45-51 dB.}

  \label{fig:average-storage-bitrate-savings}
\end{figure}

}

\newcommand{\awsBreakevenFigure}{
\begin{figure}[t]
  \centering
    \includegraphics[width=\linewidth]{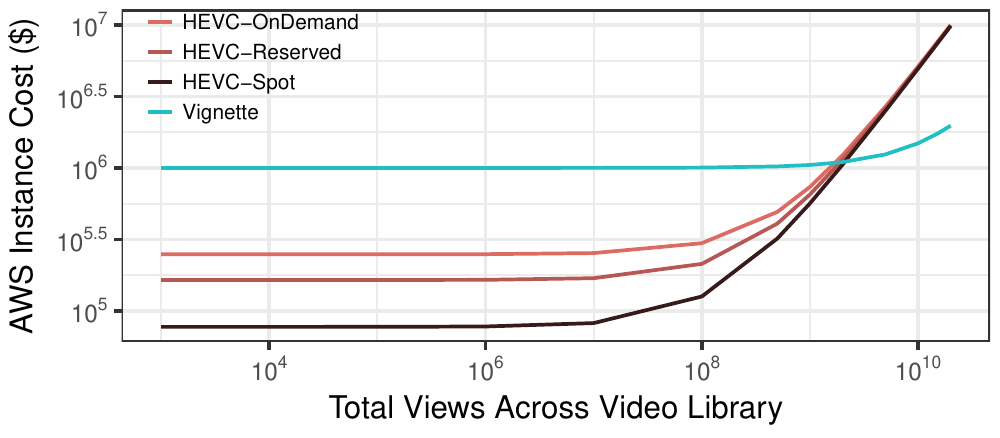}
  \caption{Estimated AWS costs for deploying \name versus traditional video transcoding. \name's additional compute cost is amortized after $\scriptstyle\sim$2 billion video views over a 1-million video library.}

  \label{fig:aws-utilization}
\end{figure}

}
\newcommand{\powerFigure}{
\begin{figure}[t]
  \centering
    \includegraphics[width=\linewidth]{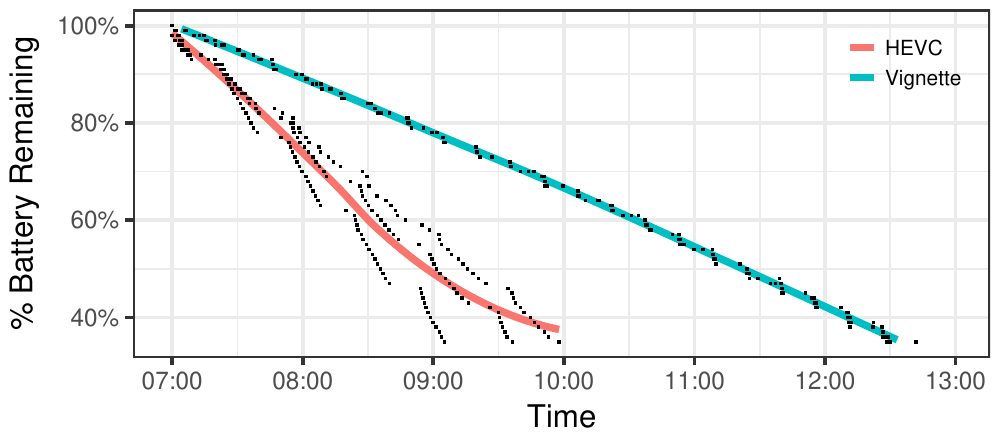}
  \caption{Time to dissipate a Google Pixel 2 phone battery from 100\% to 30\% when viewing \hevc and \name videos continuously. \name videos provide 1.67$\times$ longer video playback on mobile phones.}

  \label{fig:power}
\end{figure}

}

\begin{document}

\date{}

\title{\name: Perceptual Compression for Video Storage and Processing Systems}
%
\author{
{\rm Amrita Mazumdar, Brandon Haynes, Magdalena Balazinska, Luis Ceze, Alvin Cheung, Mark Oskin}\\
University of Washington
} 

\maketitle

\begin{abstract}

Compressed videos constitute 70\% of Internet traffic, and video upload growth rates far outpace compute and storage improvement trends.
Past work in leveraging perceptual cues like \textit{saliency}, i.e., regions where viewers focus their perceptual attention, reduces compressed video size while maintaining perceptual quality, but requires significant changes to video codecs and ignores the data management of this perceptual information.

In this paper, we propose \name, a compression technique and storage manager for perception-based video compression.
\name complements off-the-shelf compression software and hardware codec implementations.
\name's compression technique uses a neural network to predict saliency information used during transcoding, and its storage manager integrates perceptual information into the video storage system to support a perceptual compression feedback loop.
\name's saliency-based optimizations reduce storage by up to 95\% with minimal quality loss, and \name videos lead to power savings of 50\% on mobile phones during video playback.
Our results demonstrate the benefit of embedding information about the human visual system into the architecture of video storage systems.

\end{abstract}

\section{Introduction}
Compressed videos constitute 70\% of Internet traffic and are stored in hundreds of combinations of codecs, qualities, and bitrates~\cite{netflix2015encoding, cisco2017zettabyte}.
Video upload growth rates far outpace compute performance and storage production today, and will continue to accelerate~\cite{cisco2016zettabyte,fontana2018storage,vbench}.
New domains of video production---e.g., panoramic (\threesixty), stereoscopic, and light field video for virtual reality (VR)---demand higher frame rates and resolutions, as well as increased dynamic range.
Further, the prevalence of mobile devices with high-resolution cameras makes it increasingly easy for humans to capture and share video.

For decades, video codecs have exploited how humans see the world, for example, by devoting increased dynamic range to spatial features (low frequency) or colors (green) we are more likely to observe.
One such perceptual cue, \emph{saliency}, describes where in a video frame a user focuses their perceptual attention.
As video resolutions grow, e.g., \threesixty video and 8K VR displays, the salient regions of a video shrink to smaller proportion of the video frame~\cite{sitzmann2018saliency}.
Video encoders can leverage saliency by concentrating bits in more perceptually interesting visual areas.
Prior work, however, focuses only on achieving bitrate reduction or quality improvement at the cost of complicated, non-portable prototypes designed for a single codec implementation \cite{li2011visual,hadizadeh2014vidcomp,8117038,5223506}.
In this work, we address the challenges of storing and integrating this perceptual data into video storage and processing systems.

Large-scale video systems generally fall into two classes: entertainment streaming, and social media video services; saliency-based compression can provide benefits to both.
For entertainment services, which maintain small numbers of videos to be streamed at many resolutions and bitrates, saliency-based compression reduces the storage cost of maintaining many bitrates and resolution scales of these videos.
For social media services distributing a vast video library from many users, it reduces outbound network bandwidth.
As perceptual trackers, e.g., VR headsets, become more popular, saliency-based compression can use eye gaze information to further improve perceptual compression.
This feedback loop can be used to tune saliency-based compression as an initially viral video decreases in popularity, or to reduce bandwidth while streaming video to a \threesixty video player.

In this paper, we describe \name, a video storage system that leverages perceptual information to reduce video sizes and bitrates.
\name is designed to serve as a backend for large-scale video services, such as content delivery systems or social media applications.
\name has two components: a compression scheme, \textit{\nameCompress}, and a storage manager, \textit{\nameStore}.
\nameCompress leverages a new saliency-based compression algorithm to achieve up to 95\% lower bitrates while minimally reducing quality.
\nameStore uses a simple API to trigger saliency-based compression when needed, allowing applications to trade off between faster traditional compression and \name's smaller video sizes.
The system uses low-overhead metadata, can be easily integrated into existing media storage structures, and remains transparent to standard video applications.

\name is \textit{not} a new standalone codec or compression standard.
Instead, it extends existing, modern codecs to take advantage of the untapped perceptual compression potential of video content, especially high-resolution video served in VR and entertainment settings.
As a result, off-the-shelf software and hardware accelerators can decompress \name's perceptually compressed videos with no modifications.
We implement \name as an extension to \lightdb~\cite{lightdb}, a database management system for video.
Our prototype of \name demonstrates cost savings to cloud video providers and power savings during mobile video playback.

This paper makes the following contributions:

\begin{enumerate}[leftmargin=14pt]
  \item \textbf{Systems support for perceptual video compression.} We propose \name, a system for producing and managing perceptually compressed video data. \name produces videos that are 80--95\% smaller than standard videos, consume 50\% less power during playback, and demonstrate minimal perceived quality loss.
  \item \textbf{A forward-compatible encoding pipeline.} \name leverages existing features of modern video codecs to implement perceptual compression, and can be deployed in any video processing system that supports such codecs, such as \hevc or \avone. 
  \item \textbf{Custom storage for perceptual data.} \name's storage manager efficiently stores and manages perceptually compressed videos and is integrated in a modern video processing database system. \nameStore supports both a heuristic-guided search for fast perceptual compression and an exhaustive mode to compute an optimal saliency-based compression configuration.
\end{enumerate}

\noindent To our knowledge, this is the first work to consider storage management of perceptually-compressed video information.
We evaluate the limits of saliency-based compression in a video storage system with a collection of modern and high-resolution video datasets.
Using a neural network trained to predict content saliency and an off-the-shelf \hevc encoder, \name's compression scheme can reduce bitrate requirements by 80--95\%.
Our results show that \name can reduce whole-system power dissipation by 50\% on a mobile phone during video playback.
Quantitative evaluation and user study results validate that these bitrate and power savings come at no perceived loss in video quality.

\section{Background: Perceptual Compression Using Saliency Maps}

\label{sec:saliency}
Saliency is a widely-utilized measure of the perceptual importance of visual information.
Saliency data encodes the perceptual importance of information in a video, such as foreground and background or primary and secondary objects.
Video codecs already use some perceptual information, like motion and luminance, to improve compression performance~\cite{hevc}, but new modes of video viewing (such as with a VR headset) introduce the opportunity to integrate richer cues from the human visual system~\cite{lee2012perceptualcodingsurvey}.
Saliency is one such perceptual cue.
This section provides background on saliency, compares methods for generating and encoding saliency information, and introduces the machine learning technique \name uses to gather perceptual information about video data.
We also describe \emph{tiles}, the codec feature we use to compress videos with saliency information.

\noindent\textbf{Salience Maps and Detection Algorithms}: Saliency-detection algorithms highlight visually significant regions or objects in an image.
A saliency map captures visual attention in the form of a heatmap, where the map's values correspond to the salience of pixels in the input.
In this paper, we visualize saliency maps as grayscale video frames or heatmaps for clarity.

In the past, saliency information was hard to generate accurately without detailed per-video information, such as hand annotation or detailed eye gaze logs.
Moreover, the low latency and poor spatial resolution of eye-tracking devices prevented effective deployment of eye-tracker-based saliency prediction~\cite{bulling}.
VR headsets, however, are natural environments for the efficient deployment of eye tracking, and they have motivated improvements in the performance and accuracy of eye trackers~\cite{Whitmire:2016:ESC:2971763.2971771}.
Recent work combining these improved eye tracker-generated fixation maps with deep learning has improved the accuracy of saliency prediction algorithms, especially for natural images and video~\cite{bylinskii2016saliency}.

\noindent\textbf{Systems Support for Perceptual Video Compression}: Prior work has investigated many techniques for including saliency information in video compression, reducing bitrate at iso-quality by 20-60\%.
However, these techniques required significant changes to video codecs, i.e., maintaining full-frame saliency maps to use as additional input~\cite{8117038}, computing saliency models on-the-fly at high computational cost~\cite{5223506}, or solving complex optimization problems to allocate video bits~\cite{li2011visual}, as the quality of saliency map generation was the limiting factor to deploying perceptual compression.
Recent interest in applying machine learning techniques to problems in visual comprehension resulted in accurate saliency prediction models that effectively mimic the human visual system~\cite{bylinskii2016saliency}.
Moreover, interest from VR developers in deploying fast and accurate eye tracking for improved VR experiences further improved the accuracy of saliency maps with high quality fixation data, leading to a virtuous cycle of saliency map prediction and improvement~\cite{Whitmire:2016:ESC:2971763.2971771}.
The final challenge in closing the gap for deploying perceptual video compression is to design storage systems that manage and support perceptual information.

\noindent\textbf{Tiled Video Encoding}: \name uses tiles to implement perceptual compression.
Tiling a video divides a single video stream into independent regions that are encoded as separate decodable streams \cite{6547985}.
Encoders can code tiles at separate qualities or bitrates, and decoders can decode tiles in parallel.
Tiles are simple to express using standard encoding libraries, like \texttt{FFmpeg}~\cite{ffmpeg} and are supported in many video codecs.
Restricting our implementation to native tiling features introduces some loss of detail compared to designing a custom encoder.
Standard encoders only support rectangular tiles, and cannot leverage motion across tiles during encoding process.
Using only native features, however, guarantees that our compression scheme is compatible with any modern codec that implements tiling, like \hevc or \avone~\cite{avone}.
As video standards and codec efficiency improve, using general codec features to perform encoding and manage storage ensures that perceptual information remains useful.

\section{\name System Overview}

We designed \name to be easily deployed in existing video storage systems and transparent to video applications that do not require perceptual information.
\Cref{fig:high-level-architecture} shows how \name can be deployed on a typical video storage system, with \nameCompress used during the transcoding pipeline, and \nameStore managing the integration of perceptual information with video data.


\noindent\textbf{\nameCompress} uses native features found in modern video codecs.
Our implementation of \nameCompress produces videos that work out-of-the-box with any system that supports \hevc, including hardware accelerators.
\nameCompress perceptually compresses videos by enumerating configurations of video tiles and saliency-quality correspondences to maximize quality while minimizing video size.
The algorithm has three high-level steps: generate a saliency map for a given video file (\cref{subsec:supersal}), determine the optimal number of rows and columns, which we call a ``tile configuration'', to spatially partition the video into (\cref{subsec:tiles}), and select a mapping of saliency values to encoder qualities, for each tile (\cref{subsec:mapping}).

\noindent\textbf{\nameStore} manages perceptual information as simple metadata embedded within videos or maintained in the storage system.
This reduces storage complexity for data management and ensures \name data is transparent to saliency-unaware video applications such as VLC or Optasia~\cite{optasia2016lu}.
\nameStore can use a neural network to generate saliency information or collect them from end-user video viewing devices.
The storage manager supports the following features: low-overhead perceptual metadata transmitted alongside video content, without impeding the functionality of applications that choose not to use it (\cref{sec:system:metadata}), storage management policies to trigger one-time perceptual compression during ``open loop'' mode, support for refining perceptual video compression with cues from user viewing devices in a ``closed loop'' mode (\cref{sec:system:interface}), and a heuristic-based search for faster perceptual compression (\cref{subsec:search-algo}).
\architectureOverviewFigure

\section{\name Perceptual Compression Design}
\label{sec:tiled-compression}

\nameCompress uses off-the-shelf video codec features to encode perceptual information and improve coding efficiency.
Our technique takes a video as input, generates a per-frame saliency map for the video, and aggregates the per-frame maps into a single video saliency map.
\nameCompress then transcodes the input video with a tiled encoding, where the quality of each tile corresponds to the saliency of the same tile in the video's saliency map.
It uses only the native features of the \hevc codec to ensure compatibility with other video libraries.

\subsection{Automatically Generating Saliency Maps}
\label{subsec:supersal}

We use MLNet (\cite{mlnet2016}) to automatically generate a corresponding saliency map for a video input.
\cref{fig:saliency-tiles-overview} shows the saliency map generated for a video frame and how the generated maps capture the visual importance of a given video frame.
MLNet uses Keras with Theano~\cite{chollet2015keras, theano} to perform saliency prediction from video frames.
The process requires decoding the video and processing each frame through the neural network to produce output saliency maps.
We accumulate the per-frame saliency maps into a single map by collecting the maximum saliency for each pixel in the frame across the video file.
These aggregated saliency values produce a single saliency map of importance across the video.
Because video storage systems slice videos into short segments (10-20 seconds) for better coding efficiency, these video saliency maps capture aggregate saliency information without oversaturating the saliency heatmap.

\subsection{Leveraging Saliency With Tiled Video Encoding}
\label{subsec:tiles}

Once a saliency map for each video is produced, we then use it to perceptually encode videos with the tiling feature in \hevc~\cite{hevc}.
To produce saliency-based tiled video encoding, we divide a video segment spatially into tiles and then map each tile to a quality setting.
The saliency map's value at each tile determines the tile's quality setting.
For simplicity and generality, the tiling patterns we use are rectangular tiles with uniform width and height across the video frame.
We use the same tile configuration throughout the entire 10-20 second video segment for coding simplicity.
We select the number of rows and columns in each a tiling pattern based on either an exhaustive search of all tile configurations or a heuristic-guided search, described in \cref{subsec:search-algo}.

While tiling is simple and provides coding benefits, a given tile configuration can incur overheads from introducing suboptimal encoding boundaries.
Tiles are self-contained video units that can be decoded separately.
They cannot compress information beyond per-tile boundaries.
As a result, information that may be efficiently coded using partial frames in a standard encoding must be repeated if it appears in multiple tiles.
A poor tile configuration produces less efficient videos than a standard encoding pass, especially for fast-moving scenes.

We minimize the penalty of adding tile boundaries in areas that would benefit from being encoded together by exhaustively enumerating all tile configurations.
We consider only uniform-sized tiles by evaluating across all row-column pairs a video frame allows.
The \hevc standard constrains the minimum size of row and column tiles, which restricts the row-column tile configurations allowed.
In practice, we enumerate tile configurations ranging from 2$\times$2 to 5$\times$10 and 10$\times$5, compress the tiles according to their saliency values, and measure the resulting bitrate and video quality achieved.
This exhaustive enumeration takes about 30 minutes per 15-second video to find the best tile configuration with our experimental setup.

\saliencyTilesOverviewFigure

\subsection{Mapping Saliency to Video Quality Rates}
\label{subsec:mapping}

Each \hevc tile is encoded at a single quality or bitrate setting throughout the video stream, requiring \nameCompress to select per-tile encoding qualities.
We deconstruct saliency maps into per-tile parameters by mapping the highest encoding quality to the maximum saliency value in the tile's saliency map.
Selecting the video encoding quality that corresponds to a tile's saliency value is less straightforward.
Mapping saliency to video quality involves determining how video quality should be expressed during encoding and how saliency should correspond with that quality measure.

\hevc exposes different modes of controlling quality and bitrate, such as constant bitrate or constant rate factor, with varying levels of effort and efficiency.
For evaluation simplicity, we use a perceptually-controlled version of a target bitrate, where the target bitrate either corresponds to the bitrate of the original video or is specified by the API call.
The highest-saliency tiles in the video are assigned the target bitrate, and tiles with lower saliency are assigned lower bitrates, with a minimum bitrate of 10\% the original video bitrate.
As shown in ~\cref{fig:saliency-tiles-overview}, we encode a 0-255 saliency map as discrete bitrates corresponding linearly from a minimum value to the target bitrate or quality, which is the maximum.
Because \name supports standard codec features, target bitrate could be replaced with a codec's quality control, i.e. constant rate factor, as well.

\section{\nameStore System Design}
\label{sec:system}

\systemOverviewFigure

We now describe \name's storage manager for maintaining perceptual video information.
\nameStore uses low overhead metadata to encode perceptual data and a heuristic-guided search to reduce the compute load of generating perceptual transcodings.
\nameStore's metadata representation reduces full-resolution frames to a small number of bytes, and its heuristic search algorithm reduces the time taken to find an optimal tile configuration by $\sim$30$\times$ in our experiments.

\subsection{Overview of \nameStore}
\label{sec:system:goals}

\nameStore exposes perceptual video compression to applications by providing three features: (1) transparent perceptual metadata, (2) simple storage management policies, and (3) a search algorithm that reduces transcoding cost.
We embed perceptual metadata as a side channel within the video container.
Standard video containers (i.e., mp4) encapsulate saliency information along with video content, so that applications with and without perceptual support can decode \name videos.
A \threesixty video player, for example, can initialize videos to be oriented in the direction of a high-saliency region it decodes from \name metadata, but the videos can also be played traditionally in a standard video player like VLC. 

\nameStore can be used in both open and closed-feedback loops for perceptual transcoding; \Cref{fig:system-overview} shows how \nameStore can switch between an ``open loop'' mode, where video is perceptually compressed once based on automatically generated saliency maps, and a ``closed loop'' mode, where perceptually compressed video can be updated based on cues from user-end viewing devices.
The heuristic search feature included in \nameStore leverages intrinsic video features to enable $\sim$30$\times$ faster perceptual transcoding at near-optimal quality results.

\nameStore operates like similar large video management services~\cite{vbench, huang2017sve, netflix2018shotbased}.
Upon upload, it chunks videos into segments, typically 6-12 seconds in length.
Each video segment consists of one keyframe and an ensuing set of predicted frames.
\nameStore can perform perceptual compression on a per-video basis, or across the video library when a specified condition is met (e.g., low storage capacity, or video popularity decreasing beneath a threshold).

\subsection{Saliency Map Metadata}
\label{sec:system:metadata}

Video storage systems maintain containers of compressed video data that store relevant video features in metadata.
\nameStore adopts this approach, and injects a small amount ($\sim$100 bytes) of saliency metadata inside each video container.
We encode this map as a bitstring that includes fields for the number of rows and columns used for tiled saliency and the saliency weights for each tile.
These bitstrings typically range in size from 8--100 bytes.
\Cref{fig:video-metadata} shows how this metadata is included as a saliency \texttt{trak}, similar to other metadata atoms in a video container.

\videoMetadataFigure

\subsection{\nameStore API}
\label{sec:system:interface}

\begin{table*}[t]
\centering
\caption{\name API}
\label{table:policies}
\rowcolors{2}{gray!20}{white}

\resizebox{2\columnwidth}{!}{%
\begin{tabular}{lll} \toprule
Function           & Compression Type & Data required                                                                                          \\ \midrule
\texttt{transcode}        & General          & \texttt{\textless{}IN video, IN CRF/target bitrate, OUT video\textgreater{}}                                        \\
\texttt{vignette\_transcode} & Perceptual       & \texttt{\textless{}IN video, (IN CRF/target bitrate,) OUT video, OUT saliency metadata\textgreater{}}                 \\
\texttt{vignette\_squeeze}   & Perceptual       & \texttt{\textless{}IN video, IN CRF/target bitrate, OUT video\textgreater{}} \\
\texttt{vignette\_update}    & Perceptual       & \texttt{\textless{}IN video, IN fixation map, OUT video, OUT saliency metadata\textgreater{}} \\ \bottomrule
\end{tabular}
}
\end{table*}

The \nameStore API defines functions to support the open- and closed-loop modes shown in~\cref{fig:system-overview}.
\Cref{table:policies} shows the programming interface for \name, which includes three perception-specific operations: \texttt{vignette\_transcode()}, \texttt{vignette\_squeeze()}, and \texttt{vignette\_update()}.
Each API operation ingests a video and some required parameters and outputs a video with any generated perceptual metadata encapsulated in the video container.

The \name API is included as a shared library linked into~\lightdb.
System developers using \nameStore to manage video data can write storage policies or preconditions to execute \nameStore functions for a specific video or collection of videos.
For instance, a social media service could apply perceptual compression as videos decrease in popularity to reduce storage capacity.
A VR video-on-demand service that ingested eye tracking information could apply perceptual compression as new perceptual information is collected for certain videos.

\noindent\textbf{Transcode Functions.}
Transcode operations express the most basic \nameStore function, video transcoding.
When a new video is uploaded to the storage system, the storage manager triggers the general-purpose \texttt{transcode()} function to transcode the video to any specified bitrates and formats for content delivery.
This function takes as input a video and target quality parameter, expressed either by CRF or bitrate, and produces a regularly transcoded video.

The \texttt{vignette\_transcode()} function is the default saliency-based API call.
It takes as input a video and an optional quality or bitrate target, and produces both a video and its corresponding generated saliency metadata.
When \texttt{vignette\_transcode} is triggered, \nameStore generates new saliency maps, and then compresses the video according to the target quality expressed.

\nameStore's transcode functions use similar signatures, letting the system easily switch between regular and perceptual compression when storage system pressure changes.
Including saliency information as a metadata stream included in the video file container makes it transparent to saliency-agnostic applications or commands like \texttt{mediainfo} or \texttt{ffprobe}.

\noindent\textbf{Quality Modulation Functions.}
As noted in \cref{subsec:mapping}, \nameCompress maps saliency to quality levels for each tile.
A \texttt{vignette\_squeeze()} call will re-compress a video using a specified, reduced bitrate or quality threshold.
It takes in a video, target bitrate, and saliency mapping and produces the newly compressed video.
For instance, \texttt{vignette\_squeeze(input.mp4,100k)} transcodes a previously saliency-encoded video from a higher bitrate to a maximum of 100kbps in the most salient regions.
The \texttt{vignette\_squeeze()} function will recompress videos from a higher quality mapping to a lower one, but it will not transcode low-quality videos to a higher-quality mapping to avoid encoding artifacts.
This function only executes transcoding and compression with pre-generated saliency metadata, but does not update or generate new saliency metadata.
A system can invoke \texttt{vignette\_squeeze()} before video data is sent to smaller cache or in preparation for distribution to devices with smaller displays.

\noindent\textbf{Functions for Updating Perceptual Maps.}
\nameStore also supports a ``closed-loop'' mode, where saliency maps are updated with new information from eye tracking devices.
To invoke this mode, \nameStore uses the \texttt{vignette\_update()} function to ingest and re-process videos with new perceptual information.
A 2-dimensional eye tracker map is easy to convert to the saliency map input used in \nameCompress.
Similar to how \name constructs per-video saliency maps, \texttt{vignette\_update()} updates the video's saliency map with eye tracker information by executing a weighted average of the original map and the input eye tracker map.
The update function takes in a fixation map and generates a new metadata bitstream of saliency information that is attached to the video container.

\subsection{Heuristic Search for Tile Configurations}
\label{subsec:search-algo}

Most of \name's computation overhead comes from the exhaustive search over tile configurations for a given video.
This exhaustive search is typically performed once, upon video upload, but consumes significant processing time.
\nameStore contributes a lower cost search algorithm that achieves near-optimal results with a $\sim$30$\times$ performance improvement, for situations where fast saliency-based transcoding is required, e.g., for a newly uploaded video.
Depending on available resources, a video storage system could choose the exhaustive search for optimal results or heuristic-guided search for faster processing.

\name's search technique uses motion vector information from encoded video streams to estimate the size of video tiles.
It enumerates tile configurations that group regions of high motion together, and selects a configuration that minimizes the difference in motion vector values across tiles.
This heuristic approximates the observation that high-motion areas should not be divided across multiple tiles.

The algorithm extracts motion vector information from encoded videos using {{MPEGflow}}~\cite{mpegflow} and requires one transcoding pass.
Similar to our tile configuration search from \Cref{subsec:tiles}, the search exhaustively evaluates tile configurations of the motion vectors.
The search evaluates the motion encapsulated by tiles under a configuration and chooses the configuration with the minimum deviation of motion vectors in each tile.
This heuristic approximates the result of exhaustive encoding but uses much less computation.
Yet, this technique works well because good tile configurations are able to encapsulate redundant motion or frequency information with a single tile, rather than replicate it across tiles.
Compared with an exhaustive search, which can transcode a video hundreds of times to empirically produce the optimal tile configuration, our algorithm produces a result $\sim$30$\times$ faster than the exhaustive method and within 1 dB of the best-PSNR result when executed over the videos we use in our evaluation.

\section{Methodology}
\label{sec:methodology}

We implement \name by extending \lightdb~\cite{lightdb}, a database management system for VR videos.
\lightdb{} lets developers declaratively express queries over large-scale video and uses a rule-based optimizer to maximize performance.
Developers can easily express \hevc-based saliency encoding in \lightdb's query language by combining its \texttt{Encode}, \texttt{Partition}, and \texttt{Subquery} operators:
\begin{lstlisting}[style=VRQL]
Decode("rtp://...")
  >> Partition(Time, 1, Theta, $\pi$ / rows, Phi, $2\pi$ / cols)
  >> Subquery([](auto& partition) {
        return Encode(partition, $saliency\_mapping$(partition) })
  >> Store("output");
\end{lstlisting}

\noindent In this example, \texttt{Partition} divides the input video into tiles, \texttt{Encode} transcodes each tile with the corresponding \texttt{saliency\_mapping} value as an argument, and  \texttt{Subquery} executes the given operation over all the partitioned tiles.
We also wrote our object recognition queries for \cref{subsec:qos} in \lightdb to simulate video analytics workloads.
To generate saliency maps, we used MLNet~\cite{mlnet2016} with publicly-available weights trained on the SALICON~\cite{huang2015salicon}, which achieves 94\% accuracy on the MIT300 saliency benchmark.

\noindent\textbf{Baseline: } We compare \name against the HEVC encoding implementations included with \texttt{FFmpeg}.
We configure \texttt{FFmpeg} with support for NVENCODE~\cite{nvenc} GPU-based encoding of \hevc video, as it is supported by large-scale video services and devices~\cite{de2016large}.
We also implement \nameCompress on top of \texttt{FFmpeg} version \texttt{n4.1-dev}, and use the GPU-based NVENC HEVC encoder for tiled encoding.
Unless otherwise specified, we target a constrained bitrate using maximum bitrate mode (VBV) to rapidly generate results.

We performed all experiments on a single-node server running Ubuntu 16.04 and containing an Intel i7-6800K processor (3.4 Ghz, 6 cores, 15 MB cache), 32 GB DDR4 RAM at 2133 MHz, a 256 GB SSD drive (ext4 file system), and a Nvidia P5000 GPU with two discrete NVENCODE chipsets.

\noindent \textbf{Video Workload Datasets}: We use a collection of video datasets, listed in Table~\ref{table:benchmarks}, to evaluate the impact of our techniques across different classes of video.
Standard video formats and emerging VR formats comprise our evaluation datasets.
The former include representative workloads from Netflix~\cite{netflix2016data} and YouTube~\cite{vbench}.
The VR and emerging video datasets highlight demands of ultra high-definition (UHD) formats such as 360$^\circ$ video~\cite{saliency-map} and the Blender stereoscopic and UHD open source films~\cite{blender}.
To construct a representative sampling of Blender video segments, we partitioned the movies in the Blender dataset (``Elephants Dream'', ``Big Buck Bunny'', ``Sintel'', and ``Tears of Steel'') into 12-second segments, and selected five segments that covered the range of entropy rates present in each film.

\benchmarkInformationFigure

In this collection of datasets, we found that the {\tt vbench} ``desktop'' video, a 5-second computer screencast recording, responded poorly during all compression evaluations because of its low entropy and content style, so we excluded it from our evaluation results.
We discuss this style of video in relation to \name further in \cref{sec:related}.
We also replaced Netflix's single ``Big Buck Bunny'' video segment with the same video content from Blender's stereoscopic, 4K, 60 frames-per-second version of the video.

\noindent \textbf{Quantitative Quality Metrics}:
We measured video encoding quality using two quality metrics, peak signal-to-noise ratio (PSNR) and eye-weighted PSNR (EWPSNR).
PSNR reports the ratio of maximum to actual error per-pixel by computing the per-pixel mean squared error and comparing it to the maximum per-pixel error.
PSNR is popular for video encoding research, but researchers acknowledge that it fails to capture some obvious perceptual artifacts~\cite{netflix2016data}.
Acceptable PSNR values fall between 30 and 50 dB, with values above 50 dB considered to be lossless~\cite{vbench}.
For saliency prediction evaluations, researchers developed eye-weighted PSNR to more accurately represent human perception~\cite{li2011visual}.
EWPSNR prioritizes errors perceived by the human visual system rather than evaluating PSNR uniformly across a video frame.
We computed EWPSNR using the per-video saliency maps described in~\cref{sec:tiled-compression} as ground truth.

\exampleVidSaliencyFigure

\section{Evaluation}
\label{sec:eval}

We designed our evaluation to answer the following questions:

\begin{enumerate}[leftmargin=10pt,topsep=0pt,itemsep=0pt,parsep=0pt,partopsep=0pt]
 \item \textbf{Storage:} What storage and bandwidth savings does \name provide? How do tile configurations affect compression gains and quality?
 \item \textbf{Quality of Service:} How does \name's compression technique affect quality of service (QoS) of video services like video streaming (perceptual quality user study) or machine learning (speed, accuracy)? 
 \item \textbf{Compute Overhead:} What is the computational overhead of \name's compression algorithm and storage manager?
 \item \textbf{Data Center \& Mobile Cost:} How do Vignette's storage and network bandwidth savings impact video storage system and mobile viewing costs?
\end{enumerate}

\subsection{Storage and Bandwidth Savings}
\label{subsec:storage}
To evaluate the storage and bandwidth benefits of \name, we applied \nameCompress to the corpus of videos described in \Cref{sec:methodology}.
We transcoded our video library at iso-bitrate in salient regions and decreased bitrate linearly with saliency to a minimum 10\% target bitrate in the lowest saliency tiles, as illustrated in~\cref{fig:saliency-tiles-overview}.
In these experiments, we examine how our transcoding performs across a range of resolutions and workloads, as is expected in a video storage system.

\noindent \textbf{Impact of Tiling on Compression and Quality: } We first examined the impact of tiling on compression benefits using a fixed saliency map.
We used an exhaustive tile configuration search and evaluated all tile sizes to identify an optimal number of tiles for each video.
We observed that, given a fixed saliency map, optimal tile configurations to maximize storage savings and quality varied based on entropy and video content.
Some videos benefited from many small tiles, while others performed best with fewer large tiles.

\tileVsSizeFigure

The smallest tile size we evaluated were 64 pixels in breadth, but most videos performed best with tiles having a breadth of 300--400 pixels.
As \cref{plot:tile-vs-size} shows, this experiment indicated that the optimal tile configuration for a video is content-dependent and can vary from four tiles to forty, and that tile configuration is an important component of tile-based compression.

\avgStorageBitrateSavings

\noindent \textbf{Overall Compression, Bandwidth, Quality: }We next explored peak compression, bandwidth, and quality savings by applying \name to our video corpus and evaluating compression and quality savings.
We used the results of our exhaustive tile search to identify the best compression-quality configurations for each video.
\cref{fig:average-storage-bitrate-savings} shows aggregate storage savings, partitioned by  dataset.
Overall, we find that \nameCompress produces videos that are 1--15\% of the original size when maintaining the original bitrate in salient regions.
These compression savings include the fixed overhead of perceptual metadata, which is $<$100 B for all videos.
Datasets with higher video resolutions (Blender, VR-360) demonstrated the highest compression savings.
The vbench dataset, which is algorithmically chosen to have a wide variance in resolution and entropy, exhibits a commensurately large variance in storage reduction.
Of the videos with the lowest storage reduction, we find that each tends to have low entropy, large text, or other 2D graphics that are already efficiently encoded.

\Cref{table:dataset-quality} shows the average reduction in bitrate and resulting quality, measured in PSNR and EWPSNR.
Our results show that EWPSNR results are near-lossless for each benchmark dataset, while the PSNR values---which do not take the human visual processing system into account---nonetheless remain acceptable for viewing.
\Cref{fig:example-vid-saliency} highlights a \name video frame from the Netflix dataset, with an output PSNR of 36 dB and EWPSNR of 48 dB.
Overall, the results indicate that \nameCompress provides acceptable quality for its compression benefit.

\begin{table}[t]
  \caption{Average bitrate reduction and quality measurements for \nameCompress by dataset. For PSNR and EWPSNR, $>$ 30 dB is acceptable for viewing, 50 dB+ is lossless.}
  \label{table:dataset-quality}
  \centering
    \begin{tabular}{lccc}
      \toprule
       & Bitrate  & PSNR & Eye-weighted \\
      Benchmark & Reduction &   (dB) & PNSR (dB)  \\ \midrule
      vbench       &   85.6 \%     & 39       &   51                                    \\
      Netflix      & 98.6\phantom{ \%}       & 34        &  45                                 \\
      VR-360      & 98.8\phantom{ \%}        & 36        &  45                                \\
      Blender     & 98.2\phantom{ \%}        & 39        &  49               \\ \bottomrule
    \end{tabular}
\end{table}

\subsection{Quality of Service}
\label{subsec:qos}
To understand the impact of perceptual compression on common video system workloads, we evaluated quality of service (QoS) delivered by \name for two applications: entertainment streaming with a user study and evaluation of a video analytics application that performs object recognition.
These applications optimize for different QoS metrics: perceptual quality for entertainment video, and throughput and accuracy for object recognition.

\userStudyFigure


\noindent \textbf{Perceptual Quality User Study:} We ran a user study to quantify viewer perception of our saliency-based compression.
The study presented users with two versions of the same video: one encoded with \hevc at 20 Mbps, the other with \nameCompress.
The \nameCompress videos were randomly chosen to be either 1 Mbps, 5 Mbps, 10 Mbps, or 20 Mbps.
The study asked users their preference between the matched pairs for 12 videos.
The bitrate of the \nameCompress video varied randomly across the questionnaire.
The goal was to discover if viewers prefer \nameCompress to \hevc, and, if so, if those preferences are more or less pronounced at different bitrate levels for \name.

The 12 videos included three videos from each dataset, selected to cover a range of entropy levels.
Each video was encoded at a target bitrate (1Mbps, 5Mbps, 10Mbps, or 20Mbps), and the questionnaire randomly selected which bitrate to serve.
We distributed the questionnaire as a web survey and ensured videos played correctly in all browsers by losslessly re-encoding to \avc.

We recruited 35 naive participants aged 20--62 (51\% women, 49\% men) from a college campus to participate in the study.
\cref{plot:user-study} shows the results averaged across subjects and videos.
When \name videos are encoded at 1 Mbps in the most salient regions, 72\% users preferred the \hevc baseline.
However, for \name videos encoded at 5, 10, and 20 Mbps, users either could not tell the difference between \hevc and \name, or preferred \name videos 60\%, 79\%, and 81\% of the time, respectively.
This suggests that video systems can deliver \name-encoded videos at 50-75\% lower bitrate with little perceived impact.


\noindent \textbf{Object classification: } Video storage and processing systems often perform analytics and machine learning tasks on their video libraries at scale~\cite{poms2018scanner,shen2017deepvideo,zhang2017livevideoanalytics}.
To evaluate any performance degradation in latency or quality from using \nameCompress, we profile \name while running  YOLO~\cite{redmon2017yolo}, a popular fast object recognition algorithm.
We compare against baseline \hevc-encoded videos to evaluate if \name incurs any additional cost in a video processing setting.
\Cref{table:yolo} shows that using \name-compressed videos provides some speedup when decoding videos for object recognition, but this benefit is overshadowed by the cost of running YOLO.
Examining accuracy, we find that \name videos maintain 84\% accuracy on average, compared to the baseline \hevc videos.
We find that accuracy on the YOLO task is lowest for the videos in the VR360 suite, and tends to correspond to the areas where the video is distorted from the equirectangular projection.
While saliency-compressed videos can provide slight benefits for video analytics latency, especially if video decoding is the system bottleneck, future work should investigate how to optimize saliency-based compression for video analytics.

\begin{table}[t]
  \centering
  \caption{\name Speedup and Accuracy Compared to \hevc Baseline on YOLO Object Recognition.}
  \label{table:yolo}
\begin{tabular}{ccc} \toprule
  Decode              & Total Speedup             &  Average    \\
  Speedup             & (Decode + YOLO)           & Accuracy \\ \midrule
  34.6\% $\pm$ 14.3\% & 2.6\% $\pm$ 2.2\% & 84\% $\pm$ 14\% \\ \bottomrule
\end{tabular}
\end{table}

\computeTable

\subsection{Compute Overhead}
\nameCompress bears the additional processing overhead of executing a neural network to generate or update saliency maps.
\nameStore can switch between an exhaustive or more computationally-efficient heuristic tile configuration search to uncover optimal tile configurations for a video.
We benchmarked the latency of the combined saliency and transcoding pipeline in two modes: exhaustive, which generates saliency maps per frame and exhaustively evaluates tiling, and heuristic, which uses the heuristic search algorithm to select a tile configuration within 0.25 dB of the best-PSNR choice (\cref{subsec:search-algo}).
;
\Cref{table:compute} shows generating saliency maps in either mode dominates computation time for \name, and that our heuristic search is 33$\times$ faster than an exhaustive search.
This step, however, is only executed once per video and off the critical path for video streaming workloads.

\awsBreakevenFigure

\subsection{Analytical Model of \name Data Center and Mobile Costs}
\label{subsec:datacentermobile}
We use our evaluation results to model \name's system costs at scale for data center storage and end-user mobile power consumption. While these results are a first-order analysis, they suggest the potential benefit of deploying \name.

\noindent\textbf{Data center compute, storage, and network costs. }
Given the high compute cost of \name, we evaluate the break-even point for systems that store and deliver video content.
We used Amazon Web Services (AWS) prices from July 2018 in the Northern California region to characterize costs.
We use a \texttt{c5.xlarge} instance's costs for compute, S3 for storage, and vary the number of videos transferred to the Internet as a proxy for video views.
We assume a video library of 1 million videos that are 10 MB each, encoded at 100 different resolution-bitrate settings (as in~\cite{huang2017sve,netflix2018dynamicopt}) to produce $\sim$500 TB of video data.
We estimate baseline compute cost to be a two-pass encoding for each video at $\$0.212$ / sec and \name's transcode computation to be $5\times$ a baseline transcode.
Larger companies likely use Reserved or Spot Instance offerings, which provide better value for years-long reservation slots or non-immediate jobs; they are 36\% and 73\% cheaper, respectively.
For storage, we estimate costs to be \$0.023 / GB on S3 and assume Vignette-compressed videos would be 10\% of the original videos (\cref{subsec:storage}).
Transferring data out from S3 costs \$0.05 / GB; this cost is where \name achieves the majority of its savings.


\Cref{fig:aws-utilization} shows how different compute pricing models produce different savings at small numbers of video library views, but that \name becomes cost-effective at large video viewing rates.
For all compute pricing levels, a system would need to service $\sim$2 billion views across a million-video library before \name's compute overhead would be amortized across transmission and storage savings.
This number is easily reached by large video services; Facebook reported 8 billion daily views in 2016~\cite{fbviews}.

\noindent\textbf{Mobile Power Consumption.}
We explicitly designed \name to work with the \hevc{} standard so off-the-shelf software and hardware codecs could decompress \name videos.
\nameCompress's tiling strategy, however, makes video bitstream density highly non-uniform across the visual plane.
This results in inefficiency for hardware architectures that decode variably-sized tiles in parallel.
On the other hand, even such designs will achieve a higher overall power efficiency because of the reduced file sizes to decode and display.
To investigate whether \name videos can achieve power savings, we profiled power consumption on a Google Pixel 2 phone during video playback of \name videos and standard \hevc-encoded videos.

\powerFigure

We measured battery capacity on a Google Pixel 2 running Android 8.1.0, kernel 4.4.88-g3acf2d53921d, playing videos on MX Player v1.9.24 with ARMv7 NEON instructions enabled.
When possible, MX Player used hardware acceleration to decode videos.\footnote{MX Player only supported decoding stereoscopic videos with the software decoder.}
We disabled display and button backlights, as well as any configurable sensors or location monitors, to minimize extraneous power consumption.
We logged battery statistics each minute using 3C Battery Monitor Widget v3.21.8.
We conducted three trials, playing the 93-file video library in a loop until battery charge dissipated from 100\% to 30\%, for our \hevc baseline and \name videos.

\Cref{fig:power} shows our results.
We found that \name video enabled 1.6$\times$ longer video playback time with the same power consumption, or, $\sim$50\% better battery life while viewing a fixed number of videos.
While hardware decoder implementations are typically proprietary, these results indicate that perceptual compression has benefits for mobile viewers, as well as cloud video infrastructure.

\section{Related Work}
\label{sec:related}

\noindent \textbf{Saliency-based compression}: \name builds on a large body of work in saliency-based compression.
Early work improved the accuracy of saliency prediction~\cite{li2011visual, lee2012perceptualcodingsurvey}, the speed of computing saliency~\cite{GUPTA20131006, zund2013content, 5223506}, or coding efficiency~\cite{zund2013content, 8117038, 5223506,hadizadeh2014vidcomp,sitzmann2018saliency}.
These existing solutions require custom versions of outdated codecs or solving costly optimization problems during each transcoding run.
\name fundamentally differs from other contributions in perceptual compression by introducing a system design that can flexibly use \textit{any} saliency prediction algorithm or video codec, rather than narrowly focusing on accuracy, speed, or efficiency of saliency prediction.
The limitations of prior work specifically influenced \name's design as a storage manager that is compatible with existing codecs, uses low-overhead metadata, and exposes a simple API for integration.

More recently, multimedia and networking research optimized streaming bandwidth requirements for \threesixty and VR video by decreasing quality outside the VR field-of-view~\cite{Fan:2017:FPV:3083165.3083180,fov-cloud-ryoo,saliency-map,visualcloud2017haynes}; while similar in spirit to perceptual compression, this only compresses to non-visible regions of a video.
Sitzmann \etal~\cite{sitzmann2018saliency} observe the impact of leveraging saliency for VR video compression and identified key perceptual requirements, but do not address the production or distribution of saliency-compressed videos.

\noindent\textbf{Video streaming and storage systems}: The rise of video applications has driven significant recent work in processing and storage systems for video content.
Social media services like Facebook or YouTube distribute user-uploaded content from many types of video capture devices to many types of viewing devices, typically serving a small number of popular or livestreamed videos at high quality and low latency, as well as a long tail of less popular videos~\cite{fblive, tang2017popularvid}.
These workloads motivated the introduction of custom media storage infrastructure and fault-tolerant frameworks for processing video uploads at scale~\cite{beaver2010haystack, muralidhar2014f4, huang2017sve, vbench}.
Entertainment platforms like Netflix and Amazon Video have smaller amounts of video data than social media services, but incur significantly more network traffic to distribute videos broadly.
These services maintain user experience by transcoding videos at high quality for a range of heterogeneous devices and bitrate requirements, tailoring encode settings by title, streaming device, and video scene~\cite{netflix2016mobilecoding,netflix2015pertitle,netflix2018dynamicopt,netflix2018shotbased}.
For both domains, Vignette is a complementary design that solves the challenges of integrating perceptual information with video storage.

\section{Future Work and Limitations}

\noindent\textbf{Reducing compute overhead.} \name's high one-time compression cost is its biggest drawback, but can be improved. Its performance stems from the use of a highly accurate but slow neural network for saliency prediction, which does not yet use a GPU or any modern DL framework optimizations. Further, this expensive compression is run only once, and is easily amortized across many views (\cref{subsec:datacentermobile}).


\noindent\textbf{Saliency for screencasts and 2D graphics.} We eliminated one outlier video, because the saliency model performed poorly.
Incorporating recent saliency models specifically designed for 2D visualizations~\cite{bylinskii2017vizsaliency} would likely resolve the issue.

\noindent\textbf{Integration with other video system optimizations.} We could further improve \name by building on other optimizations that work with off-the-shelf video standards.
For instance, \name's heuristic search algorithm could include power and performance information from open-source video transcoding ASICs~\cite{asicclouds, zhang2017racetosleep} to target more power-efficient tiling configurations.
VideoCoreCluster~\cite{liu2016greenvid} demonstrated energy-efficient adaptive bitrate streaming in real-time using a cluster of low-cost transcoding ASICs, which \name could leverage for better server transcoding performance.
Using Fouladi~\etal's parallel cloud transcoding could also improve \name's transcode latency, and \name's saliency-based tiling could integrate with their codesigned network transport protocol and video codec to better tune streaming quality~\cite{fouladi2018salsify, fouladi2017excamera}.
At the physical storage layer, Jevdjic \etal's approximate video storage framework, which maps video streams to different layers of error correction, could be coupled with \name's saliency mapping for more aggressive approximation of non-salient video regions~\cite{jevdjic2017approxvid}.
Integrating \name with these systems could further improve power efficiency during playback, transcoding latency, or archival video storage durability.

\section{Conclusion}
Video data continues to grow with increased video capture and consumption trends, but leveraging perceptual cues can help manage this data.
This paper proposes integrating perceptual compression techniques with video storage infrastructure to improve storage capacity and video bitrates while maintaining perceptual quality.
\name combines automatic generation of perceptual information with a video transcoding pipeline to enable large-scale perceptual compression with minimal data overhead.
Our storage system supports a feedback loop of perceptual compression, including updates as an application gathers data from sources such as eye trackers.
Our offline compression techniques deliver storage savings of up to 95\%, and user trials confirm no perceptual quality loss for \name videos 50-75\% smaller in size.

\name's design complements the contributions of existing large-scale video storage and processing systems.
Video systems can use \name to further improve storage capacity or in anticipation of video workloads that produce perceptual information.
As VR video consumption and new perceptual markers --- such as eye trackers in VR headsets --- grow in popularity, \name's techniques will be critical in integrating perceptual compression at large scale for higher quality, lower bitrate video. 

\bibliographystyle{plain}
\bibliography{\jobname}

\end{document}